\documentclass[aps,prd,preprint,eqsecnum,showpacs]{revtex4}
\usepackage{latexsym}
\usepackage{graphicx}% Include figure files
\begin{document}
% \draft

\thispagestyle{empty}

{\baselineskip0pt
\leftline{\large\baselineskip16pt\sl\vbox to0pt{\hbox{\it Department of
Physics}
               \hbox{\it Osaka City  University}\vss}}
\rightline{\large\baselineskip16pt\rm\vbox to20pt{\hbox{OCU-PHYS-224}
            \hbox{AP-GR-22}\hbox{YITP-05-4}
\vss}}%
}
\vskip3cm
% number in the upper right hand corner of the title page in preprint mode.
% Use the 'preprint numbers' class option to override journal defaults
%\preprint{}

\title{Gravitational Radiation from Cylindrical Naked Singularity}
\author{Ken-ichi Nakao$^{1}$\footnote{Electronic
address: knakao@sci.osaka-cu.ac.jp}
and 
Yoshiyuki Morisawa$^{2}$\footnote{Electronic
address: morisawa@yukawa.kyoto-u.ac.jp}}
\affiliation{
$^{1}$Department of Physics, Graduate School of Science, 
Osaka City University, Osaka 558-8585, Japan\\
$^{2}$Yukawa Institute for Theoretical Physics, Kyoto University, 
Kyoto 606-8502, Japan}
\date{\today}

\begin{abstract}                % DON'T CHANGE THIS LINE
We construct an approximate solution which describes 
the gravitational emission from a naked singularity formed 
by the gravitational collapse of a cylindrical thick shell 
composed of dust. The assumed situation is that the collapsing 
speed of the dust is very large. In this situation, the metric 
variables are obtained approximately by a kind of linear 
perturbation analysis in the background Morgan solution which 
describes the motion of cylindrical null dust. 
The most important problem in this study is 
what boundary conditions for metric and matter variables should be 
imposed at the naked singularity. 
We find a boundary condition that all the metric and matter 
variables are everywhere finite at least up to the first order 
approximation. This implies that the spacetime singularity 
formed by this high-speed dust collapse is very similar to that 
formed by the null dust and thus the gravitational emission from a 
naked singularity formed by the cylindrical dust collapse can be gentle.  

\end{abstract}


\maketitle
%%---------------------------%
\section{Introduction}

The cosmic censorship hypothesis\cite{Ref:penrose69} 
is a crucial ansatz for the theorems of black 
holes\cite{Ref:hawking72}. There are 
two versions for this hypothesis. For spacetimes with 
physically reasonable matter fields, the weak version 
claims that a spacetime singularity as a result of generic 
non-singular initial data is not visible from infinity, 
while the strong version claims that a 
spacetime singularity developed from non-singular initial 
data is invisible for any observer. A singularity censored 
by the strong version is called a naked singularity, 
while a singularity censored by the weak version is called 
a globally naked singularity. However, previous theoretical 
studies have revealed several candidates for the counter-examples 
of this hypothesis\cite{Ref:Joshi-text,Ref:NAKED1,Ref:NAKED2,Ref:NAKED3}. 
Although further detailed and careful studies about these candidates 
are necessary, these might have some physical importance. 

If a globally naked singularity forms, what can we detect from 
there? In connection with this issue, Nakamura, Shibata and 
one of the present authors (KN) have proposed a conjecture; 
if globally naked singularities form, 
large spacetime curvatures in the neighborhood of 
these can propagate away to infinity in the form of gravitational
radiation, and as a result, almost all of the mass of these naked 
singularities is lost through this large gravitational 
emission\cite{Ref:DRIES-UP}. 
If this conjecture is true, the 
formation processes of the globally naked singularities 
might be very important for gravitational-wave 
astronomy\cite{Ref:GW-US,Ref:GW-J}. The cylindrically symmetric 
system will play an important role in getting significant 
information about this issue by the following reasons; 
in the asymptotically flat case, the singularities 
are necessarily naked\cite{Ref:HOOP,Ref:Hayward};  
there is a degree of gravitational radiation; this 
system is very simple. There are a few numerical studies on  
gravitational emissions by cylindrically symmetric gravitational 
collapse\cite{Ref:Piran,Ref:Dust-Shell,Ref:Chiba}. 

Recently, the present authors have investigated the 
high-speed collapse of cylindrically symmetric thick shell composed 
of dust\cite{Ref:paperI} and perfect fluid with non-vanishing 
pressure\cite{Ref:paperII}. In these studies, we use the 
high-speed approximation which is 
a kind of the perturbation analysis in the background Morgan 
solution describing the motion of a cylindrical null dust. 
In the first paper, it has been revealed that in the 
gravitational collapse of a cylindrically symmetric thick 
dust shell, the thinner shell leads to the larger amount 
of gravitational radiation.  This result gives a resolution 
of an apparent inconsistency between the previous works. 
In the second paper, we showed that the pressure decelerates the 
collapsing velocity so significantly that the high-speed 
approximation scheme breaks down before a singularity forms 
if the equation of state is moderately hard. For example, 
in the case of mono-atomic ideal gas which seems to be physically 
reasonable in the high-energy state due to the asymptotic 
freedom of elementary interactions, the high-speed collapse 
is prevented by the pressure and thus there is a possibility that 
the pressure halts the formation of a spacetime singularity 
in the case of this cylindrical gravitational collapse. 
However, we should note that if initial collapsing velocity is 
very large, a region with very large mass concentration can be realized 
in the neighborhood of the symmetric axis $r=0$. 
In this region, the tidal force for free 
falling observers can be so large that general relativity 
breaks down and the quantum theory of gravity will be necessary 
to understand the physical processes realized there. 
Such a region can be regarded as a singularity for 
general relativity, called a `spacetime border'\cite{Ref:border}. 
By contrast, in case that the equation of state is sufficiently soft, 
the high-speed collapse is maintained until a globally 
naked singularity forms.

In this paper, we investigate the gravitational waves from 
a naked singularity formed by the high-speed gravitational 
collapse of a cylindrical thick shell composed of dust. 
In the previous paper\cite{Ref:paperI}, 
we have investigated the generation process of those 
in the causal past of the Cauchy horizon associated with 
the naked singularity, i.e., the region 
which does not suffer the influence of the naked singularity 
(for details of the Cauchy horizon, causal future and 
causal past, see, for example, 
the textbook by Wald\cite{Ref:Wald-text}). 
On the other hand, in this paper, we focus on the generation 
process of the gravitational waves in the causal future 
of the naked singularity. 
In order to know what happens in the causal future of 
the naked singularity, we need specify the boundary condition 
at the naked singularity. This is equivalent to defining 
the naked singularity as a physical entity. We would like to show that 
in the high-speed approximation scheme, there is a boundary condition 
that all the metric and matter variables are everywhere finite 
at least up to the first order. There are several studies on how to 
fix the boundary condition for test fields or gravitational
perturbations in static naked 
singular spacetimes\cite{Ref:Wald,Ref:HM,Ref:IH,Ref:IW,Ref:GHI}. 
By contrast, the present case is dynamical, 
and as far as we are aware, this is the first example of gravitational 
emissions from a naked singularity developed from non-singular 
initial data. 

This paper is organized as follows. In Sec.II, we present
the basic equations of the cylindrically symmetric dust system.
In Sec.III, we derive the basic equations in 
the high-speed approximation scheme which describe the 
gravitational collapse with velocity almost 
equal to the speed of light $c$. In Sec.IV, we consider the 
boundary condition at the background naked singularity and then 
present solutions for the basic equations given in Sec.III.
In Sec.V, we study the gravitational 
emissions from the naked singularity. Finally Sec.VI is 
devoted to summary and discussion. 

In this paper, we adopt $c=1$ unit and basically follow the 
convention of the Riemann and metric tensors and the notation 
in the textbook by Wald\cite{Ref:Wald-text}.

\section{Cylindrically Symmetric Dust System}

In this paper, we focus on the spacetime 
with whole-cylinder symmetry which is defined by 
the following line element\cite{Ref:C-energy},
\begin{equation}
ds^{2}=e^{2(\gamma-\psi)}\left(-dt^{2}+dr^{2}\right)
+e^{-2\psi}R^{2}d\varphi^{2}+e^{2\psi}dz^{2}.
\end{equation}
Then Einstein equations are
\begin{eqnarray}
&&\gamma'=\left({R'}^{2}-{\dot R}^{2}\right)^{-1}
\biggl\{
RR'\left({\dot \psi}^{2}+{\psi'}^{2}\right)
-2R{\dot R}{\dot \psi}\psi'
+R'R''-{\dot R}{\dot R}' \nonumber \\ 
&&~~~~-8\pi G\sqrt{-g}\left(R'T^{t}{}_{t}+{\dot R}T^{r}{}_{t}\right)
\biggr\}, 
\label{eq:einstein-1} \\
&&{\dot \gamma}=-\left({R'}^{2}-{\dot R}^{2}\right)^{-1}
\biggl\{
R{\dot R}\left({\dot \psi}^{2}+{\psi'}^{2}\right)
-2RR'{\dot \psi}\psi' 
+{\dot R}R''-R'{\dot R}' \nonumber \\
&&~~~~-8\pi G\sqrt{-g}\left({\dot R}T^{t}{}_{t}+R'T^{r}{}_{t}\right)
\biggr\}, \\
&&{\ddot \gamma}-\gamma''={\psi'}^{2}-{\dot \psi}^{2}-{8\pi G\over R}
\sqrt{-g}T^{\varphi}{}_{\varphi}, \\
&&{\ddot R}-R''
=-8\pi G\sqrt{-g}\left(T^{t}{}_{t}+T^{r}{}_{r}\right), \\
&&{\ddot \psi}+{{\dot R}\over R}{\dot \psi}-\psi''
-{R'\over R}\psi' 
=-{4\pi G\over R}\sqrt{-g}\left(T^{t}{}_{t}+T^{r}{}_{r}-T^{z}{}_{z}
+T^{\varphi}{}_{\varphi}\right),
\label{eq:einstein-2}
\end{eqnarray}
where a dot means the derivative with respect to $t$ while a prime means 
the derivative with respect to $r$. 

As mentioned, we consider the dust fluid as a matter field. 
The stress-energy tensor is 
\begin{equation}
T_{\mu\nu}=\rho u_{\mu}u_{\nu},
\end{equation}
where $\rho$ is rest mass density 
and $u^{\mu}$ is a 4-velocity of the fluid element.  
Due to the assumption of the whole-cylinder symmetry, 
the components of the 4-velocity $u^{\mu}$ are written as
\begin{equation}
u^{\mu}=u^{t}\left(1,-1+V,~0,~0\right).
\end{equation}
By the normalization $u^{\mu}u_{\mu}=-1$, $u^{t}$ is expressed as 
\begin{equation}
u^{t}={e^{-\gamma+\psi}\over \sqrt{V\left(2-V\right)}}.
\end{equation}

Here denoting the determinant of the metric tensor $g_{\mu\nu}$ 
by $g$, we introduce new density variable $D$ defined by
\begin{equation}
D:={\sqrt{-g}\rho u^{t}\over \sqrt{V\left(2-V\right)}}
={R e^{\gamma-\psi}\rho \over V\left(2-V\right)}.
\label{eq:D-def}
\end{equation}
The components of the stress-energy tensor are then expressed as
\begin{eqnarray}
\sqrt{-g}T^{t}{}_{t}&=&-e^{\gamma-\psi}D, 
\label{eq:st-tensor-tt}\\
\sqrt{-g}T^{r}{}_{t}&=&e^{\gamma-\psi}D(1-V)=-\sqrt{-g}T^{t}{}_{r}, \\
\sqrt{-g}T^{r}{}_{r}&=&e^{\gamma-\psi}\left(1-V\right)^{2}D,
\label{eq:st-tensor-rr}
\end{eqnarray}
and the other components vanish. 

The equation of motion $\nabla_{\alpha}T^{\alpha}{}_{\beta}=0$ leads to 
\begin{eqnarray}
\partial_{u}D&=&-{1\over2}(DV)'
+{D\over2}(1-V)\left\{2\partial_{u}(\psi-\gamma)
-V({\dot \psi}-{\dot\gamma})\right\}, \label{eq:conservation-full}\\
&& \nonumber \\
D\partial_{u}V&=&(1-V)\partial_{u}D+{1\over2}\left\{V(1-V)D\right\}'
-{D\over2}\left\{2\partial_{u}(\psi-\gamma)-V({\dot \psi}-{\dot\gamma})\right\}, 
  \label{eq:Euler-full}
\end{eqnarray}
where $u=t-r$ is the retarded time and $\partial_{u}$ is the partial 
derivative of $u$ with fixed advanced time $v=t+r$. The first equation
comes from the $t$-component while the second one comes from the
$r$-component. The $z$- and $\varphi$-components are trivial.  

To estimate the energy flux of the gravitational radiation, 
we investigate $C$-energy $E$ and its flux vector proposed by 
Thorne\cite{Ref:C-energy}. $C$-energy $E=E(t,r)$ is the energy 
per unit coordinate length along $z$-direction 
within the radius $r$ at time $t$, which is defined by 
\begin{equation}
E={1\over8}\left\{1+e^{-2\gamma}\left({\dot R}^2-{R'}^2\right)\right\}.
\end{equation}
The energy flux vector $J^\mu$ associated to the C-energy is defined by
\begin{equation}
\sqrt{-g}J^\mu=\left(
{\partial E\over \partial r},-{\partial E\over \partial t},
0,0\right).
\end{equation}
By its definition, $J^{\mu}$ is divergence free. Using the equations of 
motion for the metric variables, we obtain the following 
expression of $C$-energy flux vector, 
\begin{eqnarray}
\sqrt{-g}J^{t}&=&{e^{-2\gamma}\over 8\pi G}
\biggl\{RR'({\dot \psi}^{2}+{\psi'}^{2}) 
-2R{\dot R}{\dot \psi}\psi'
-8\pi G\sqrt{-g}(R'T^{t}{}_{t}+{\dot R}T^{r}{}_{t})\biggr\}, \\
\sqrt{-g}J^{r}
&=&{e^{-2\gamma}\over 8\pi G}
\biggl\{R{\dot R}({\dot \psi}^{2}+{\psi'}^{2}) 
-2RR'{\dot \psi}\psi' 
-8\pi G\sqrt{-g}(R'T^{r}{}_{t}-{\dot R}T^{r}{}_{r})\biggr\},
\end{eqnarray}
and the other components vanish.

\section{High Speed Approximation Scheme}

Let us consider the ingoing null limit of the cylindrical dust  
fluid. From Eqs.(\ref{eq:st-tensor-tt})-(\ref{eq:st-tensor-rr}), 
the stress-energy tensor is written in the following form,  
\begin{equation}
T_{\mu\nu}={e^{3(\psi-\gamma)}D\over R}k_{\mu}k_{\nu},
\label{eq:st-tensor}
\end{equation}
where
\begin{equation}
k^{\mu}=(1,-1+V,~0,~0).
\end{equation}
The timelike vector $k^{\mu}$ becomes the ingoing null vector 
in the limit of $V\rightarrow0_+$.  
Hence in the limit of $V\rightarrow0_+$ with $D$ fixed, 
the stress-energy tensor agrees with that of the collapsing 
null dust. This means that in the case of very large 
collapsing velocity, i.e., $0<V\ll1$, the dust fluid 
system will be well approximated by the null dust system. 
Then we treat the ``deviation $V$ of the 4-velocity from null'' 
as a perturbation and perform linear perturbation analyses. 

In the case of the complete null dust $V=0$, the solution is 
easily obtained as 
\begin{eqnarray}
\psi&=&0, \\
\gamma&=&\gamma_{\rm B}(v), \label{eq:gamma-BG}\\
R&=&r, \label{eq:R-BG} \\
8\pi G De^{\gamma}&=&{d\gamma_{\rm B}\over dv}, \label{eq:D-BG}
\end{eqnarray}
where $\gamma_{\rm B}(v)$ is an arbitrary function of the 
advanced time $v$. 
This solution was first obtained by Morgan\cite{Ref:Morgan} 
and was studied subsequently by Letelier and Wang\cite{Ref:LW} 
and Nolan\cite{Ref:Nolan} in detail. 
Assuming the null energy condition, the density variable 
$D$ is non-negative. This means that $\gamma_{\rm B}$ is non-decreasing 
function of $v$. 

The situation of the collapsing null dust solution 
can be understood by Fig.1. In this paper, 
we assume that the density variable $D$ has a compact support 
$0<v<v_{\rm w}$ which is depicted by a shaded region in Fig.1. 
We can find in Eqs.(\ref{eq:st-tensor}) and 
(\ref{eq:R-BG}) that if $D$ does not vanish at the symmetric axis 
$r=0$, components of the stress-energy tensor $T_{\mu\nu}$ with respect 
to the coordinate basis diverge there, and the same is true for 
the Ricci tensor by Einstein equations. 
This is a naked singularity which is depicted 
by the dashed line at $r=0$ in $0< t< v_{\rm w}$ in Fig.1. 
Although all the scalar polynomials of Riemann tensor vanish there, 
freely falling observers suffer infinite tidal force at $r=0$ in 
$0<t<v_{\rm w}$; this type of singularity is called {\it p.p.} curvature 
singularity\cite{Ref:HE}. The Cauchy horizon associated 
with this intermediate naked singularity is represented 
by the short dashed line $t=r$ in Fig.1. On the other hand, 
the region at $r=0$ in $t\geq v_{\rm w}$ is conical singularity 
which is depicted by the thick dotted line in this figure. 

\begin{figure}
{\resizebox{10cm}{!}{\includegraphics{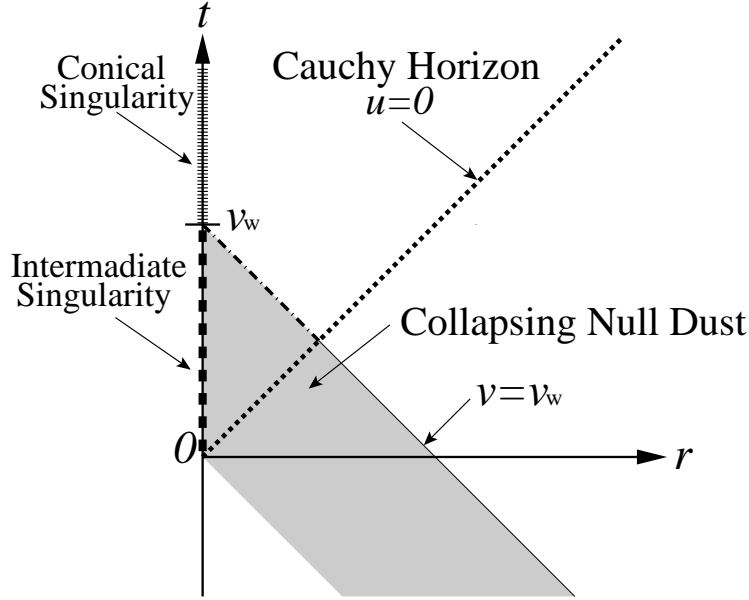}}}
%        \centerline{\epsfxsize 10cm \epsfysize 10cm \epsfbox{morgan.eps}}
\caption{Morgan's cylindrical null dust solution. 
There is the null dust in the shaded region; $D(v)>0$ for 
$0<v<v_{\rm w}$. The dashed line on $t$-axis corresponds to 
the intermediate singularity at which an observer suffers infinite 
tidal force although any scalar polynomials of the Riemann tensor 
do not vanish.  On the other hand, the dotted line on $t$-axis 
is the conical singularity. The Cauchy horizon is short dashed line of $t=r$. 
We estimate $C$-energy transfered through the null hypersurface 
$v=v_{\rm w}$ from $u=0$ to $u=v_{\rm w}$, which is depicted 
by dot-dashed line. 
}
\label{fg:morgan}
\end{figure}

We introduce a small parameter $\epsilon$ and assume the order of 
the variables as $V=O(\epsilon)$ and $\psi=O(\epsilon)$.
Further we rewrite the variables $\gamma$, $R$ and $D$ as
\begin{eqnarray}
e^{\gamma}&=&e^{\gamma_{\rm B}}(1+\delta_{\gamma}), \\
R&=&r(1+\delta_{R}), \\
D&=&D_{\rm B}(1+\delta_{D}),
\end{eqnarray}
and assume that $\delta_{\gamma}$, $\delta_{R}$ 
and $\delta_{D}$ are $O(\epsilon)$, where
\begin{equation}
D_{\rm B}:={1\over 8\pi Ge^{\gamma_{\rm B}}}{d\gamma_{\rm B}\over dv}.
\label{eq:DB-def}
\end{equation}
We call the perturbative analysis with respect to this small 
parameter $\epsilon$ the high speed approximation scheme. 

The 1st order equations with respect to $\epsilon$ are 
given as follows; the Einstein equations 
(\ref{eq:einstein-1})-(\ref{eq:einstein-2}) lead to
\begin{eqnarray}
&&\delta_{\gamma}{}'=
8\pi GD_{\rm B}e^{\gamma_{\rm B}}\left\{\delta_{\gamma}-\psi+\delta_{D} 
-2\partial_{v}(r\delta_{R})\right\} 
+(r\delta_{R})'',
\label{eq:del-r-gamma}\\
&&{\dot \delta}_{\gamma}=
8\pi GD_{\rm B}e^{\gamma_{\rm B}}\left\{\delta_{\gamma}-\psi+\delta_{D}
-2\partial_{v}(r\delta_{R})-V\right\} +(r{\dot\delta_{R}})', 
\label{eq:del-t-gamma}\\
&&{\ddot\delta}_{\gamma}-\delta_{\gamma}{}''=0, 
\label{eq:gamma-linear}\\
&&r{\ddot\delta}_{R}-(r\delta_{R})''
=16\pi Ge^{\gamma_{\rm B}}D_{\rm B}V, 
\label{eq:R-linear}\\
&&{\ddot\psi}-\psi''-{1\over r}\psi'={8\pi G \over r}
e^{\gamma_{\rm B}}D_{\rm B}V; 
\label{eq:psi-linear}
\end{eqnarray}
the conservation law (\ref{eq:conservation-full}) leads to  
\begin{equation}
\partial_{u}\left(\delta_{D}+\delta_{\gamma}-\psi\right)
=-{1\over 2D_{\rm B}}{dD_{\rm B}\over dv}V 
-{1\over2}\left(V'-{d\gamma_{\rm B}\over dv}V\right);  
\label{eq:conservation-linear}
\end{equation}
the Euler equation (\ref{eq:Euler-full}) becomes as
\begin{equation}
\partial_{u}V=0,  \label{eq:V-linear}
\end{equation}
where we have used Eq.(\ref{eq:conservation-linear}). 

$C$-energy $E$ up to the first order of $\epsilon$ is given by
\begin{equation}
E={1\over8}\left[1-e^{-2\gamma_{\rm B}}+2e^{-2\gamma_{\rm B}}
\left\{\delta_\gamma-\left(r\delta_R\right)'\right\}\right].
\label{eq:C-energy-linear}
\end{equation}
From Eq.(\ref{eq:D-BG}), we can easily see that 
$\gamma_{\rm B}$ is constant in the vacuum region $D_{\rm B}=0$. 
Further, from Eqs.(\ref{eq:del-r-gamma}) and (\ref{eq:del-t-gamma}), 
we find that $\delta_\gamma-(r\delta_R)'$ is also constant in the 
vacuum region. Therefore up to the first order, $C$-energy 
is constant in the vacuum region. This means that in the 
vacuum region, $C$-energy flux vector $J^\mu$ vanishes up to 
the first order and thus it is a second order quantity which is given by
\begin{eqnarray}
\sqrt{-g}J^{t}&=&{r\over 8\pi G}\left({\dot \psi}^{2}+{\psi'}^{2}\right), \\
\sqrt{-g}J^{r}&=&-{r\over 4\pi G}{\dot \psi}\psi'. 
\label{eq:C-flux-linear}
\end{eqnarray}
$C$-energy flux vector takes very similar form to that of 
massless Klein-Gordon field. 

\section{Solutions for Dust Collapse}

In this section, we study 
the behaviors of perturbation variables in the causal 
future of the background naked singularity. 
Einstein equations are hyperbolic differential equations 
which send out the information of the background 
naked singularity into its causal future. Therefore 
in order to get the solutions of the basic equations for the 
perturbation variables in the causal future of the 
background naked singularity, we need specify 
the boundary condition for these at the background naked 
singularity. 

In general, it is very difficult to control the solutions 
by the boundary conditions due to the nonlinearity of Einstein 
equations. However, fortunately, there is a great advantage in 
the present situation; the perturbation variables can be regarded 
as test fields with sources in the fixed background Morgan spacetime. 
This seems to be similar situation to the problems 
investigated by Wald and a several researchers, 
i.e., the behavior of test fields in naked singular 
static spacetimes\cite{Ref:Wald,Ref:HM,Ref:IH,Ref:IW,Ref:GHI}.
However, there is a significant difference 
between the previous studies and the present case. 
In the previous cases, the effect of the naked singularity appears 
in the differential operator in the equations of motion for the test 
fields. On the other hand, in the present case, the manifestation of 
the naked singularity appears as source terms in the differential 
equations for the perturbation variables. Therefore we have to 
determine the boundary condition from different point of view from 
previous works. 

We do not know the law of physics at the spacetime singularity, 
i.e., the quantum theory of gravity and hence, at present, 
it is difficult to impose a physically satisfactory 
boundary condition at the naked singularity. 
There is a possibility that the quantum 
gravity puts the additional source terms on the 
right hand side of the differential equations 
(\ref{eq:del-r-gamma})-(\ref{eq:psi-linear}) 
at the background naked singularity; 
the additional source term might be a delta-function 
like one causing gravitational wave bursts, or by contrast, 
might completely cancel out the original source term 
and, as a result, lead no gravitational radiation from 
the naked singularity. 
However, at first glance, it is expected that the geometrical 
structure of the spacetime investigated here is not so 
different from the background Morgan 
spacetime even in the neighborhood of the background 
naked singularity. Thus in order to search for such a solution, 
we require that the perturbation 
variables are expressed in the form of the Maclaurin series around the 
naked singularity. As will be shown below, this condition 
guarantees that all the perturbation variables are everywhere 
finite up to the first order, and this is just what we expect.  
The cylindrical thick shell composed of dust also forms a 
naked singularity at the same 
coordinate position as the background naked singularity.  
The Cauchy horizon associated with this naked singularity 
is located also at the same coordinate position as the 
background one. 

\subsection{Solution for $\delta_\gamma$}

Equation (\ref{eq:gamma-linear}) is easily solved. In this equation, 
there is no singularity even at the background naked singularity. 
Before the appearance of the background naked singularity, i.e.,  
$t\leq0$, we impose a regularity condition to guarantee the locally 
Minkowskian  nature at $r=0$; $\delta_\gamma$ is $C^{2}$ function 
with ${\delta_\gamma}'|_{r=0}=0$. We also impose the same 
condition at $r=0$ after the background naked 
singularity formation $t>0$. 
Then we obtain general solutions for $\delta_\gamma$
\begin{equation}
\delta_\gamma=\Delta_\gamma(t-r)+\Delta_\gamma(t+r),
\end{equation}
where $\Delta_\gamma$ is an arbitrary $C^2$ function. Here we 
should note that after the appearance of the naked 
singularity $t\geq0$, the regularity 
condition ${\delta_\gamma}'|_{r=0}=0$ does no longer guarantee the 
locally Minkowskian nature since the background spacetime itself 
does not have the locally Minkowskian nature at the naked singularity. 

\subsection{Solution for $V$}

The solution for the Euler equation (\ref{eq:V-linear}) is given by
\begin{equation}
V=C_V(v),  \label{eq:Dust-V-sol}
\end{equation}
where $C_V$ is an arbitrary function which is everywhere
finite. Therefore even at the background naked singularity, 
the velocity perturbation $V$ does not vanish. This fact means 
that the type of the singularity formed at the symmetric center 
$r=0$ is changed from that of the background Morgan spacetime. 
Up to the first order of $\epsilon$, the Ricci scalar is obtained as 
\begin{equation}
R^\mu{}_\mu =-8\pi GT^\mu{}_\mu
={16\pi G e^{-\gamma_{\rm B}}D_{\rm B} V\over r}
\end{equation}
We can easily find in the above equation that 
the Ricci scalar diverges at the symmetric center $r=0$ 
if $D_{\rm B}$ and $V$ do not vanish there. 
Thus the background naked {\it p.p.} curvature singularity 
becomes the {\it s.p.} curvature singularity which is defined as 
the spacetime singularity accompanied 
by the divergence of a scalar polynomial constructed 
from the Riemann tensor\cite{Ref:HE}.   

\subsection{Solution for $\delta_R$}

Using the solution (\ref{eq:Dust-V-sol}), we can get the solution 
for $\delta_R$. For notational simplicity, we introduce a 
function $S$ defined by
\begin{equation}
S(v)=8\pi G e^{\gamma_{\rm B}}D_{\rm B}C_V.  \label{eq:source}
\end{equation}
Note that two times $S(v)$ is the source term in the equation
(\ref{eq:R-linear}) for $\delta_R$. 

As mentioned, we search for the solution 
which can be expressed by Maclaurin series in the neighborhood 
of the background naked singularity as 
\begin{equation}
\delta_R=\delta_R^{(0)}(t)+\delta_R^{(1)}(t) r
 + \delta_R^{(2)}(t)r^2+\delta_R^{(3)}(t)r^3...
\end{equation}
The source function $S$ is also assumed to 
be written by Maclaurin series as
\begin{equation}
S(v)=S^{(0)}(t)+S^{(1)}(t)r+S^{(2)}(t)r^2+....
\label{eq:s-series}
\end{equation}
Substituting the above expressions into Eq.(\ref{eq:R-linear}), 
we obtain the following equations 
\begin{equation}
-2\delta_R^{(1)}=2S^{(0)},~~~~
{\ddot\delta}_R^{(0)}-6\delta_R^{(2)}=2S^{(1)}, ~~~~
{\ddot\delta}_R^{(1)}-12\delta_R^{(3)}=2S^{(1)}, ...
\end{equation}
Thus the boundary condition at $r=0$ is given by 
\begin{equation}
{\delta_R}'|_{r=0}=-S(t).
\label{eq:R-BC}
\end{equation}
Since the expansion coefficients $S^{(i)}(t)$ of the source function 
are finite, the coefficients $\delta_R^{(i)}$ are also finite even 
in the neighborhood of the naked singularity. 

The solution for $\delta_R$ with the boundary condition 
(\ref{eq:R-BC}) is easily obtained by constructing a Green function. 
Since $r \delta_R$ should vanish at $r=0$, the Green function 
$G_R$ for $r \delta_R$ should also vanish there. Therefore 
$G_R$ is given by 
\begin{equation}
G_R(t,r;\tau,x)={1\over\pi}\int_0^\infty {dk\over k}
\sin(kr)\sin(kx)\sin\{k(t-\tau)\}.
\end{equation}
Thus the solution for $\delta_R$ is given by 
\begin{eqnarray}
\delta_R(t,r)&=&{2\over\pi r}\int_{t_{\rm i}}^td\tau
\int_0^\infty dx G_R(t,r;\tau,x) S(\tau+x) \nonumber \\
&+&{1\over r}\left\{\Delta_R(t+r)-\Delta_R(t-r)\right\},
\label{eq:Dust-delR-sol}
\end{eqnarray}
where $t_{\rm i}$ is some constant, $\Delta_R$ is an arbitrary
$C^2$ function. 

\subsection{Solution for $\psi$}

Before the appearance of the background naked singularity 
$t\leq0$, we impose the regularity condition on $\psi$ 
on the symmetric axis $r=0$; $\psi$ is $C^2$ function satisfying 
$\psi'|_{r=0}=0$. 
However, at the naked singularity $0 < t < v_{\rm w}$, 
this condition is not satisfied because the right hand side of 
Eq.(\ref{eq:psi-linear}) diverges at $r=0$ if $S|_{r=0}\neq0$. 
Even in such a situation, we can find solutions everywhere finite. 

Here we assume that $\psi$ can be written in the form of the 
Maclaurin series 
around $r=0$, 
\begin{equation}
\psi=\psi^{(0)}(t)+\psi^{(1)}(t)r+\psi^{(2)}(t)r^2+....
\end{equation}
Substituting this expression and Eq.(\ref{eq:s-series}) 
into Eq.(\ref{eq:psi-linear}), we find 
\begin{equation}
-\psi^{(1)}=S^{(0)},~~{\ddot \psi}^{(0)}-4\psi^{(2)}=S^{(1)}
,~~{\ddot\psi}^{(1)}-9\psi^{(3)}=S^{(2)}....
\end{equation}
Thus the boundary condition at the background naked singularity 
is given by  
\begin{equation}
\psi'|_{r=0}=-S(t). \label{eq:B-condition}
\end{equation}
The above condition guarantees that the solution 
for $\psi$ is everywhere finite. 
To get the numerical values of $\psi$, we need numerical 
calculations. We will do it in Sec.V.

Here we should note that the background conical singularity 
at $r=0$ and $t\geq v_{\rm w}$ remains conical one up to the 
first order. This is because the perturbation variables $\delta_R$, 
$\delta_\gamma$ and $\psi$ behave as those of the regular 
vacuum spacetime at the background conical singularity by the 
present boundary condition.  

\subsection{Solution for $\delta_D$}

It is easy to integrate Eq.(\ref{eq:conservation-linear}). 
Using Eqs.(\ref{eq:Dust-V-sol}) and (\ref{eq:source}), 
the solution is obtained as
\begin{equation}
\delta_D=\psi-\delta_\gamma+{u\over2}
\left(2S-C_V{d\ln S\over dv}\right) +C_D(v),
\end{equation}
where $C_D$ is an arbitrary function which is everywhere 
finite.

\section{Gravitational Radiation from Naked Singularity}

In this section, we study the energy carried 
by gravitational radiation emitted from the naked singularity. 
Since the energy flux in the vacuum region is given 
by Eq.(\ref{eq:C-flux-linear}), we numerically calculate 
$\psi$ in the causal future of the background naked singularity. 
In order to do so, we adopt the retarded and advanced 
coordinates $u$ and $v$ instead of $t$ and $r$. 
Then Eq.(\ref{eq:psi-linear}) is rewritten in the form
\begin{eqnarray}
{\partial \psi \over \partial u}&=&D_u(u,v), \label{eq:wave-psi}\\
{\partial D_v(u,v) \over \partial u}&=&
{\partial D_u(u,v) \over \partial v}=
{1\over2(v-u)}
\left\{D_v(u,v)-D_u(u,v)+S(v)\right\}, 
\label{eq:wave-Dv-Du}
\end{eqnarray}

We consider the one-parameter family of the background 
null dust solutions. We adopt the thickness 
$v_{\rm w}$ of the shell as the parameter and 
set all the members in this family having the same total 
$C$-energy $E_{\rm B}$. 
Since the total $C$-energy $E_{\rm B}$ in the 
background spacetime is given by 
\begin{equation}
E_{\rm B}={1\over8}\left(1-e^{-2\gamma_\infty}\right),
\end{equation}
where $\gamma_\infty$ is the asymptotic value of $\gamma_{\rm B}$ 
for $v\rightarrow\infty$. Thus independence of $C$-energy $E_{\rm B}$ 
on the thickness $v_{\rm w}$ means that $\gamma_\infty$ 
does not depend on $v_{\rm w}$. We can see in the above equation that 
$E_{\rm B}$ is less than $1/8$. If $E_{\rm B}$ is equal to 1/8, 
$\gamma_{\infty}$ is infinite, and then the space is 
closing up in the $r$-direction. 

For notational convenience, we introduce the 
dimensionless null coordinates defined by   
\begin{equation}
x={u\over v_{\rm w}}~~~~~{\rm and}~~~~~y={v\over v_{\rm w}}.
\end{equation}
Then the asymptotic value $\gamma_\infty$ is written as
\begin{equation}
\gamma_\infty=\int_0^{v_{\rm w}} dv{d\gamma_{\rm B} \over dv}
=\int_0^1dyv_{\rm w}{d\gamma_{\rm B}\over dv}.
\label{eq:gamma-inf}
\end{equation}
Since $\gamma_\infty$ does not depend on the thickness 
$v_{\rm w}$ of the dust shell, $d\gamma_{\rm B}/dv$ should be written 
in the following form,
\begin{equation}
{d\gamma_{\rm B}\over dv}
={1\over v_{\rm w}}F_{\gamma}(y). 
\end{equation}

We set the velocity perturbation $V$ being 
\begin{equation}
V=C_V(v)=F_V(y), 
\end{equation}
so that the maximal value of $V$ does not depend on the thickness 
$v_{\rm w}$. Then we find 
\begin{equation}
S(v)=V{d\gamma_{\rm B}\over dv}={1\over v_{\rm w}}
F_{\gamma}F_{V}
=:{1\over v_{\rm w}}F_S(y).
\end{equation}

We introduce new dimensionless variables defined by 
\begin{equation}
D_x(x,y)=v_{\rm w}D_u(u,v)
~~~~~{\rm and}~~~~~
D_y(x,y)=v_{\rm w}D_v(u,v).
\end{equation}
Then Eqs.(\ref{eq:wave-psi}) and (\ref{eq:wave-Dv-Du}) take the 
following forms,
\begin{eqnarray}
{\partial \psi \over \partial x}&=&D_x, 
\label{eq:wave-psi-2}\\
{\partial D_y \over \partial x}&=&
{\partial D_x \over \partial y}=
{1\over2(y-x)}
\left(D_y-D_x+F_S\right), 
\label{eq:wave-Dv-Du-2}
\end{eqnarray}
Since $F_S$ is the function of $y$, the thickness $v_{\rm w}$ 
does disappear in the above equation. This means that 
$\psi$, $D_x$ and $D_y$ are the functions of $x$ and $y$ but 
do not depend on the thickness $v_{\rm w}$ as long as we use 
the variables $x$ and $y$ as independent variables.

\subsection{$C$-energy}

We calculate $C$-energy carried by the gravitational 
radiation from the naked {\it s.p.} curvature singularity 
formed at $r=0$. The background singularity in 
$t \geq v_{\rm w}$ is a conical 
singularity at which the source function $S$ vanishes. 
This means that $\psi$ is not generated there. 
Thus we focus on the gravitational 
radiation generated at the intermediate 
naked singularity $0<t<v_{\rm w}$. 
For this purpose, it is sufficient to investigate $C$-energy 
flux through the null hypersurface of 
$v=v_{\rm w}$ from $u=0$ to $u=v_{\rm w}$ (see Fig.1). 
Using Eq.(\ref{eq:C-flux-linear}), $C$-energy 
$\Delta E_{\rm NS}$ transfered from the intermediate background 
{\it p.p.} curvature singularity through this null 
hypersurface is given as 
\begin{equation} 
\Delta E_{\rm NS}
= {1\over 4G}\int_{0}^{v_{\rm w}}
du (v_{\rm w}-u)D_u{}^2(u,v_{\rm w})
= {1\over 4G}\int_{0}^{1}
dx (1-x)D_x{}^2(x,1). 
\label{eq:emitted-energy}
\end{equation}
From the above equation, we find that $\Delta E_{\rm NS}$ does not 
depend on the thickness $v_{\rm w}$ of the dust shell. This fact 
is different from $C$-energy carried by gravitational waves 
generated in the causal past of the Cauchy horizon\cite{Ref:paperI} 
(see also Appendix A). 

From Eq.(\ref{eq:gamma-inf}), we see that $\gamma_\infty$ is 
proportional to $F_\gamma$. 
Since the mean value $V_{\rm m}$ of the velocity perturbation $V$ 
is given by 
\begin{equation}
V_{\rm m}={1\over v_{\rm w}}\int_0^{v_{\rm w}}dvV(v)
=\int_0^1 dy F_V(y), 
\label{eq:V-mean}
\end{equation}
$V_{\rm m}$ is proportional to $F_V$. 
This means that $D_x$ is proportional to $\gamma_\infty V_{\rm m}$ 
since the source term $F_S$ is the product of $F_\gamma$ and $F_V$. 
Rewriting $\gamma_\infty$ by the background total $C$-energy as
\begin{equation}
\gamma_\infty =-{1\over2}\ln(1-8E_{\rm B}),
\end{equation}
we obtain
\begin{equation}
\Delta E_{\rm NS}={\rm Const} \times 
\left\{V_{\rm m} \ln(1-8E_{\rm B})\right\}^2.
\label{eq:dependence}
\end{equation}
Since the absolute value of the source term $|F_{\rm S}|$ 
should be much smaller than unity so that 
the high-speed approximation is valid, 
$|V_{\rm m} \ln(1-8E_{\rm B})|$ should
be much less than unity. However at present, we can not 
say whether $\Delta E_{\rm NS}$ is large compared to the total 
$C$-energy $E_{\rm B}$ or not. We need focus on a 
specific model of our interest and then calculate `Const' in 
Eq.(\ref{eq:dependence}).  

\subsection{Example}

Here we consider the following example,
\begin{eqnarray}
S(v)&=&-{18\over v_{\rm w}}V_{\rm m}\ln(1-8E_{\rm B})
y^2(1-y)^2\Theta(y)\Theta(1-y), \\
V(v)&=&6V_{\rm m}y(1-y)\Theta(y)\Theta(1-y),
\end{eqnarray}
where $\Theta$ is Heaviside's step function. 
Note that the above $S$ and $V$ have been chosen so that 
$E_{\rm B}$ and $V_{\rm m}$ are the total $C$-energy of the background
spacetime and the mean value of $V$, respectively. 
We solve Eqs.(\ref{eq:wave-psi-2}) and (\ref{eq:wave-Dv-Du-2}) 
numerically and then estimate Eq.(\ref{eq:emitted-energy}). 

We set the $C$-energy of the present system being the same as 
that of the background. By Eq.(\ref{eq:C-energy-linear}), this 
condition leads to the following condition on $u=u_{\rm i}$, 
\begin{equation}
\delta_\gamma-\left(r\delta_R\right)'=0.
\label{eq:E-condition}
\end{equation}
Eqs.(\ref{eq:del-r-gamma}) and (\ref{eq:del-t-gamma}) lead to 
\begin{equation}
\partial_v\left\{\delta_\gamma-\left(r\delta_R\right)'\right\}
=8\pi Ge^{\gamma_{\rm B}}D_{\rm B}\left\{\delta_\gamma
-\psi+\delta_D-2\partial_v(r\delta_R)
-{V\over2}\right\}.
\end{equation}
Therefore in the region of $D_{\rm B}>0$, 
the following equation should hold,
\begin{equation}
\delta_\gamma
-\psi
+\delta_D
-2\partial_v(r\delta_R)
-{V\over2}=0.
\end{equation}
Here we adopt the following initial data so that 
Eq.(\ref{eq:E-condition}) is satisfied, 
\begin{equation}
\psi=\delta_\gamma=\delta_R=0~~{\rm and}~~\delta_D
={V\over2}~~~~{\rm on}~~u=u_{\rm i}.
\end{equation}

It is not so difficult to numerically solve the 
differential equations (\ref{eq:wave-psi-2}) and (\ref{eq:wave-Dv-Du-2}). 
We set the initial condition $\psi=0$ and thus $D_y=0$ at the Cauchy
horizon $x=0$. This initial condition guarantees that there is no
ingoing flux across the Cauchy horizon and hence we can see the
gravitational radiation generated just at the naked singularity.  
The result of the three-digit accuracy is obtained as
\begin{equation}
\Delta E_{\rm NS} = 0.0643\times \left\{V_{\rm m}
\ln(1-8E_{\rm B})\right\}^2.
\end{equation}
Here for notational convenience, we introduce a new 
parameter defined by 
\begin{equation}
\varepsilon = -V_{\rm m}\ln(1-8E_{\rm B}).
\end{equation}
Since the source function $S$ is proportional to this parameter 
$\varepsilon$, $\varepsilon$ should be much less than unity 
so that the high speed approximation is applicable.  
$E_{\rm B}$ is expressed by $\varepsilon$ and $V_{\rm m}$ as
$E_{\rm B}=\left(1-e^{-\varepsilon/V_{\rm m}}\right)/8$ and thus 
we obtain  
\begin{equation}
{\Delta E_{\rm NS}\over E_{\rm B}} 
= 0.514 \times 
{\varepsilon^2 \over 1-e^{-\varepsilon/V_{\rm m}}}. 
\end{equation}
We can easily see that the above quantity is monotonically 
increasing function of both $\varepsilon$ and 
$V_{\rm m}$ for $\varepsilon>0$ and $V_{\rm m}>0$. 
In the parameter space $(\varepsilon, V_{\rm m})$, 
$\varepsilon=V_{\rm m}=1$ is the marginal point at which 
the high speed approximation must break down.  
The upper bound on $\Delta E_{\rm NS}/E_{\rm B}$ for the high-speed 
collapse is realized at this point,
\begin{equation} 
{\rm max}\left({\Delta E_{\rm NS}\over E_{\rm B}} \right)
= 0.813.
\end{equation}
The above value is rather large. The large amount of the 
gravitational radiation seems to be emitted if the collapsing 
speed $1-V_{\rm m}$ is not so large and the background 
$C$-energy $E_{\rm B}$ is not so small. 
On the other hand, in the case of very small $\varepsilon$, 
the emitted energy is 
\begin{equation}
{\Delta E_{\rm NS}\over E_{\rm B}} \simeq
0.514\times \varepsilon V_{\rm m}.
\end{equation}
Thus the emitted energy will be at most a few percents of the 
total energy of this system in the situation that the present 
approximation scheme describes the system with sufficient accuracy. 

\section{Summary and Discussion}

In this paper, we studied the gravitational radiation 
from the naked singularity formed by high-speed gravitational 
collapse of a cylindrical thick shell composed of dust. 
In this issue, it is very important what boundary 
condition for metric and matter variables 
should be imposed at the background naked singularity. 
Here we have adopted the boundary condition that 
all the perturbation variables are expressed in the form of the 
Maclaurin series at the background naked singularity. 
Then we have found a boundary condition that guarantees all 
the metric and matter variables are finite everywhere 
at least up to the first order approximation. 
At first glance, it is expected that the spacetime of 
the high-speed dust collapse is not so different from 
the background Morgan spacetime even in the neighborhood 
of the background naked singularity. The boundary condition 
adopted in the present paper realize what we expected. 
Further this implies that the high-speed approximation scheme 
is everywhere valid. 

We investigated $C$-energy sent out by gravitational 
emission from the naked singularity. In the situation that the present 
approximation scheme is valid, every perturbation variable is 
small and as a result, the emitted energy of gravitational radiation is 
small. We assumed a simple 
model and found that the upper bound of the emitted energy by the
high-speed collapse is almost the same as the total $C$-energy of the 
system. However, in the case that the high-speed approximation describes 
the system with sufficient accuracy, the emitted energy might be 
at most a few percents of the total $C$-energy of 
the system. Thus the emission of gravitational waves from cylindrical 
naked singularity can be gentle process. This result seems to imply that
the conjecture by Nakamura, Shibata and KN is
not necessarily valid. However the dust might be unrealistic
assumption and hence further investigation is necessary. 
On the other hand, the rather large value of the 
upper bound on the energy of gravitational radiation implies 
that there remains possibility of the large gravitational radiation by 
the gravitational collapse of a cylindrical thick dust shell with not so 
large collapsing velocity. This is also a future problem. 

\section*{Acknowledgements}

We are grateful to colleagues in the astrophysics and 
gravity group of Osaka City University for helpful discussion and criticism.  
KN thanks A.~Hosoya and A.~Ishibashi for useful discussion. 
This work is supported by the Grant-in-Aid for Scientific Research 
(No.16540264) from JSPS and also by the Grant-in-Aid for 
the 21st Century COE ``Center for Diversity 
and Universality in Physics'' from the Ministry 
of Education, Culture, Sports, Science and 
Technology (MEXT) of Japan.  

\appendix
\section{} %Empty argument \section{} yields `Appendix'. 

Here we consider the energy sent out by the gravitational 
radiation generated in the causal past of the Cauchy horizon 
$u=0$. The emitted energy $\Delta E_{\rm R}$ from $u=u_{\rm i}(<0)$  
to $u=0$ is given by
\begin{equation} 
\Delta E_{\rm R}(v)
= {1\over 4G}\int_{u_{\rm i}}^{0}
du (v-u)D_u{}^2(u,v)
= {1\over 4G}\int_{x_{\rm i}}^{0}
dx (y-x)D_x{}^2(x,y),
\label{eq:emitted-energy-regular}
\end{equation}
where $x_{\rm i}=u_{\rm i}/v_{\rm w}$ and 
$y=v/v_{\rm w}$ is larger than unity so that the 
estimation of $\Delta E_{\rm R}$ is done outside of the shell.  
In contrast to the case of the gravitational waves from 
the naked singularity in Sec.V, the emitted energy 
$\Delta E_{\rm R}$ depends on the thickness $v_{\rm w}$ 
of the shell. In the thin shell limit, 
$v_{\rm w}\rightarrow 0$, the region of integration 
in Eq.(\ref{eq:emitted-energy-regular}) becomes 
$(-\infty,0]$ since $x_{\rm i}\rightarrow-\infty$. 
Thus if $D_x(x,y)$ does not vanish 
sufficiently rapidly in the limit of 
$x\rightarrow -\infty$, this integral diverges. 

Let us consider the initial data of which $\psi=0$ 
at $u=u_{\rm i}$ (thus $D_v=0$), or equivalently $\psi=D_y=0$ 
at $x=x_{\rm i}$. This initial condition guarantees that there is no 
ingoing energy flux across $x=x_i$. Here we assume 
$x_{\rm i}\ll-1$ and then consider the case of $|x|\gg y$. 
In this case, Eq.(\ref{eq:wave-Dv-Du-2}) becomes as
\begin{equation}
{\partial D_x \over \partial y}\sim 
-{1\over2x}\left(D_y-D_x+F_{\rm S}\right).
\end{equation}
Substituting the expression 
\begin{equation}
D_x=A(x,y)e^{y\over2x}
\end{equation}
into the above equation, we obtain
\begin{equation}
{\partial A\over \partial y}\sim -{1\over 2x}
\left(D_y +F_{\rm S}\right)
e^{-{y\over 2x}}
\sim -{1\over 2x}
\left({\partial\psi\over\partial y} +F_{\rm S}\right).
\end{equation}
Integrating the above equation with respect to $y$, we obtain 
\begin{equation}
A\sim-{1\over2x}\left(\psi+\int_0^yF_{\rm S}(z)dz\right),
\end{equation}
where the integration constant has been set so that 
$A$ vanishes in the vacuum region inside the shell $y<0$ at 
$x=x_{\rm i}$. 
Therefore we find for $x\sim x_{\rm i}$ and $y>1$,  
\begin{equation}
D_x \sim -{M \over x},  
\end{equation}
where 
\begin{equation}
M={1\over2}\int_0^1F_S(z)dz
\end{equation}
is a positive constant. 

The above result means that if we set $x_{\rm i}$ 
being $-\infty$, $D_x(x,y)\rightarrow -M/x$ for $x\rightarrow 
-\infty$. Hence introducing some constant $\lambda\ll-1$, 
we see that in the large $x_{\rm i}$ limit,  
\begin{eqnarray}
\Delta E_{\rm R}
&=& {1\over 4G}\int_{x_{\rm i}}^{0}
dx (y-x)D_x{}^2(x,y) \nonumber \\
&=& {1\over 4G}
\left(\int_{x_{\rm i}}^{\lambda}+\int_{\lambda}^0\right)
dx (y-x)D_x{}^2(x,y) \nonumber \\
&\simeq&{1\over 4G}
\int_{x_{\rm i}}^{\lambda}
dx {M^2\over x} 
+F(y;\lambda)
\longrightarrow \infty ~~~{\rm for}
~~x_{\rm i}\longrightarrow-\infty,
\end{eqnarray}
where $F(y;\lambda)$ is some function of $y$ and 
the parameter $\lambda$.
This results means that in the thin-shell limit with 
the total $C$-energy fixed, the emitted energy 
diverges. This is consistent with the result in 
Ref.\cite{Ref:paperI}.

%
%\section{Second Appendix}

\end{document}